\documentstyle[twoside,epsf,psfig]{article}

\catcode`\@=11
\long\def\@makefntext#1{
\protect\noindent \hbox to 3.2pt {\hskip-.9pt  
$^{{\eightrm\@thefnmark}}$\hfil}#1\hfill}               %CAN BE USED 

\def\thefootnote{\fnsymbol{footnote}}
\def\@makefnmark{\hbox to 0pt{$^{\@thefnmark}$\hss}}    %ORIGINAL 
        
\def\ps@myheadings{\let\@mkboth\@gobbletwo
\def\@oddhead{\hbox{}
\rightmark\hfil\eightrm\thepage}   
\def\@oddfoot{}\def\@evenhead{\eightrm\thepage\hfil
\leftmark\hbox{}}\def\@evenfoot{}
\def\sectionmark##1{}\def\subsectionmark##1{}}

\oddsidemargin=\evensidemargin
\renewcommand{\thefootnote}{\fnsymbol{footnote}}

\newcounter{sectionc}\newcounter{subsectionc}\newcounter{subsubsectionc}
\renewcommand{\section}[1] {\vspace{12pt}\addtocounter{sectionc}{1} 
\setcounter{subsectionc}{0}\setcounter{subsubsectionc}{0}\noindent 
        {\tenbf\thesectionc. #1}\par\vspace{5pt}}
\renewcommand{\subsection}[1] {\vspace{12pt}\addtocounter{subsectionc}{1} 
\setcounter{subsubsectionc}{0}\noindent 
{\bf\thesectionc.\thesubsectionc. {\kern1pt \bfit #1}}\par\vspace{5pt}}
\renewcommand{\subsubsection}[1] {\vspace{12pt}\addtocounter{subsubsectionc}{1}
        \noindent{\tenrm\thesectionc.\thesubsectionc.\thesubsubsectionc.
        {\kern1pt \tenit #1}}\par\vspace{5pt}}
\newcommand{\nonumsection}[1] {\vspace{12pt}\noindent{\tenbf #1}
        \par\vspace{5pt}}

\newcounter{appendixc}
\newcounter{subappendixc}[appendixc]
\newcounter{subsubappendixc}[subappendixc]
\renewcommand{\thesubappendixc}{\Alph{appendixc}.\arabic{subappendixc}}
\renewcommand{\thesubsubappendixc}
        {\Alph{appendixc}.\arabic{subappendixc}.\arabic{subsubappendixc}}

\renewcommand{\appendix}[1] {\vspace{12pt}
        \refstepcounter{appendixc}
        \setcounter{figure}{0}
        \setcounter{table}{0}
        \setcounter{lemma}{0}
        \setcounter{theorem}{0}
        \setcounter{corollary}{0}
        \setcounter{definition}{0}
        \setcounter{equation}{0}
        \renewcommand{\thefigure}{\Alph{appendixc}.\arabic{figure}}
        \renewcommand{\thetable}{\Alph{appendixc}.\arabic{table}}
        \renewcommand{\theappendixc}{\Alph{appendixc}}
        \renewcommand{\thelemma}{\Alph{appendixc}.\arabic{lemma}}
        \renewcommand{\thetheorem}{\Alph{appendixc}.\arabic{theorem}}
        \renewcommand{\thedefinition}{\Alph{appendixc}.\arabic{definition}}
        \renewcommand{\thecorollary}{\Alph{appendixc}.\arabic{corollary}}
        \renewcommand{\theequation}{\Alph{appendixc}.\arabic{equation}}
        \noindent{\tenbf Appendix \theappendixc #1}\par\vspace{5pt}}
\newcommand{\subappendix}[1] {\vspace{12pt}
        \refstepcounter{subappendixc}
        \noindent{\bf Appendix \thesubappendixc. {\kern1pt \bfit #1}}
        \par\vspace{5pt}}
\newcommand{\subsubappendix}[1] {\vspace{12pt}
        \refstepcounter{subsubappendixc}
        \noindent{\rm Appendix \thesubsubappendixc. {\kern1pt \tenit #1}}
        \par\vspace{5pt}}

\newcommand{\textlineskip}{\baselineskip=13pt}
\newcommand{\smalllineskip}{\baselineskip=10pt}

\newcommand{\copyrightheading}[1]
        {\vspace*{-2.5cm}\smalllineskip{\flushleft
        {\footnotesize International Journal of Modern Physics D, #1}\\
        {\footnotesize \copyright\kern2pt World Scientific Publishing
         Company}\\
         }}

\newcommand{\publisher}[2]{{\begin{center}\footnotesize\smalllineskip 
        Received #1\\
        Revised #2
        \end{center}
        }}

\def\abstracts#1#2#3{{
        \centering{\begin{minipage}{4.5in}\footnotesize\baselineskip=10pt
        \parindent=0pt #1\par 
        \parindent=15pt #2\par
        \parindent=15pt #3
        \end{minipage}}\par}} 

\def\keywords#1{{
        \centering{\begin{minipage}{4.5in}\footnotesize\baselineskip=10pt
        {\footnotesize\it Keywords}\/: #1
         \end{minipage}}\par}}

\renewenvironment{thebibliography}[1]
        {\frenchspacing
         \ninerm\baselineskip=11pt
         \begin{list}{\arabic{enumi}.}
        {\usecounter{enumi}\setlength{\parsep}{0pt}     
         \setlength{\leftmargin 12.7pt}{\rightmargin 0pt}%FOR 1--9 ITEMS
         \setlength{\itemsep}{0pt} \settowidth
        {\labelwidth}{#1.}\sloppy}}{\end{list}}

\newcounter{itemlistc}
\newcounter{romanlistc}
\newcounter{alphlistc}
\newcounter{arabiclistc}

\newcommand{\fcaption}[1]{
        \refstepcounter{figure}
        \setbox\@tempboxa = \hbox{\footnotesize Fig.~\thefigure. #1}
        \ifdim \wd\@tempboxa > 5in
           {\begin{center}
        \parbox{5in}{\footnotesize\smalllineskip Fig.~\thefigure. #1}
            \end{center}}
        \else
             {\begin{center}
             {\footnotesize Fig.~\thefigure. #1}
              \end{center}}
        \fi}

\newcommand{\tcaption}[1]{
        \refstepcounter{table}
        \setbox\@tempboxa = \hbox{\footnotesize Table~\thetable. #1}
        \ifdim \wd\@tempboxa > 5in
           {\begin{center}
        \parbox{5in}{\footnotesize\smalllineskip Table~\thetable. #1}
            \end{center}}
        \else
             {\begin{center}
             {\footnotesize Table~\thetable. #1}
              \end{center}}
        \fi}

\def\@citex[#1]#2{\if@filesw\immediate\write\@auxout
        {\string\citation{#2}}\fi
\def\@citea{}\@cite{\@for\@citeb:=#2\do
        {\@citea\def\@citea{,}\@ifundefined
        {b@\@citeb}{{\bf ?}\@warning
        {Citation `\@citeb' on page \thepage \space undefined}}
        {\csname b@\@citeb\endcsname}}}{#1}}

\newif\if@cghi
\def\cite{\@cghitrue\@ifnextchar [{\@tempswatrue
        \@citex}{\@tempswafalse\@citex[]}}
\def\citelow{\@cghifalse\@ifnextchar [{\@tempswatrue
        \@citex}{\@tempswafalse\@citex[]}}
\def\@cite#1#2{{$\null^{#1}$\if@tempswa\typeout
        {IJCGA warning: optional citation argument 
        ignored: `#2'} \fi}}

\def\pmb#1{\setbox0=\hbox{#1}
        \kern-.025em\copy0\kern-\wd0
        \kern.05em\copy0\kern-\wd0
        \kern-.025em\raise.0433em\box0}

\def\fnt#1#2{\footnotetext{\kern-.3em
        {$^{\mbox{\scriptsize #1}}$}{#2}}}

\def\fpage#1{\begingroup
\voffset=.3in
\thispagestyle{empty}\begin{table}[b]\centerline{\footnotesize #1}
        \end{table}\endgroup}

\def\runninghead#1#2{\pagestyle{myheadings}
\markboth{{\protect\footnotesize\it{\quad #1}}\hfill}
{\hfill{\protect\footnotesize\it{#2\quad}}}}
\font\tenrm=cmr10
\font\tenit=cmti10 
\font\tenbf=cmbx10
\font\bfit=cmbxti10 at 10pt
\font\ninerm=cmr9

\font\eightrm=cmr8

\def\qed{\hbox{${\vcenter{\vbox{                  %HOLLOW SQUARE
   \hrule height 0.4pt\hbox{\vrule width 0.4pt height 6pt
   \kern5pt\vrule width 0.4pt}\hrule height 0.4pt}}}$}}

\renewcommand{\thefootnote}{\fnsymbol{footnote}}  %USE SYMBOLIC FOOTNOTE

\begin{document}
\setlength{\textheight}{7.7truein}    %FOR 2ND PAGE ONWARDS

\runninghead{Adaptive filtering techniques for interferometric data
  $\ldots$} {Adaptive filtering techniques for interferometric
  data$\ldots$}

\normalsize\textlineskip
\thispagestyle{empty}
\setcounter{page}{1}

\copyrightheading{}             %{Vol.~0, No.~0 (1999) 000--000}

\vspace*{0.88truein}

\fpage{1}
\centerline{\bf Adaptive filtering techniques for interferometric data 
preparation~:}
\vspace*{0.035truein}
\centerline{\bf removal of long-term sinusoidal signals and oscillatory transients}
\vspace*{0.37truein}
\centerline{\footnotesize E. CHASSANDE-MOTTIN (1) and S. DHURANDHAR (1) \& (2)}
\vspace*{0.015truein}
\centerline{\footnotesize\it (1) Albert-Einstein-Institut, Am
M\"uhlenberg, 1}
\vspace*{0.015truein}
\centerline{\footnotesize\it (2) IUCAA, Postbag 4, Ganeshkhind}
\baselineskip=10pt
\centerline{\footnotesize\it Pune 411 007, India}
\vspace*{0.225truein}
\publisher{(received date)}{(revised date)}

\vspace*{0.21truein} \abstracts{We propose an adaptive denoising
  scheme for poorly modeled non-Gaussian features in the gravitational
  wave interferometric data. Preliminary tests on real data show
  encouraging results.}{}{}

\vspace*{10pt} \keywords{Gravitational wave interferometric data,
  detector characterization, transient denoising, LMS method, adaptive
  line enhancement}

%\textlineskip                  %) USE THIS MEASUREMENT WHEN THERE IS
%\vspace*{12pt}                 %) NO SECTION HEADING

\setcounter{footnote}{0}
\renewcommand{\thefootnote}{\alph{footnote}}

\vspace*{1pt}\textlineskip      %) USE THIS MEASUREMENT WHEN THERE IS
\section{Motivation}    %) A SECTION HEADING
\vspace*{-0.5pt}
\noindent
Several large-scale interferometric gravitational wave detectors will
come on-line soon, such as LIGO in the U.S., the French/Italian Virgo
project, GEO600 the German/British interferometer and TAMA in
Japan. Gravitational wave detectors produce an enormous volume of
output. Data analysis techniques will have to be developed to
optimally extract the weak signature of a gravitational wave from
these data. Many of the techniques developed so far are based on
matched filtering and assume stationary Gaussian noise.

However, the real data stream from the detectors is not expected to
satisfy the stationary and Gaussian assumptions. This disparity
between standard Gaussian assumptions and real data characteristics
poses a major problem to the direct application of matched filtering
techniques in particular for burst sources such as black hole binary
quasinormal ringings~\cite{creighton97:_listen} or inspiral waveforms~
\cite{allen99:_obser_galax}. In fact, the data from the Caltech 40
meter proto-type interferometer has the expected broadband noise
spectrum, but superposed on this are several other noise features:
such as long-term sinusoidal disturbances coming from suspensions and
electric main harmonics and also ringdown transients occurring
occasionally, typically due to servo-controls instabilities or
mechanical relaxation in suspension system etc.  While no precise
 model can be given for this noise until the detector
is completed and fully tested, matched filtering techniques cannot be
used to locate/remove these noisy signals.

We propose a denoising method based on \textit{adaptive linear
  prediction} techniques which does not require any precise \textit{a
  priori} information about the noise characteristics. Although our method does not
pretend to optimality, we believe that its simplicity makes it useful
for data preparation and for the understanding of the first data.

In the following, we present the structure of the proposed algorithm
and some results obtained with the data from the Caltech 40 meter
proto-type interferometer~\cite{abramovici96:_improv_ligo}. For a
more detailed presentation, we refer the reader to~
\cite{chassande-mottin00:_adapt}.

\section{Adaptive linear prediction}
\vspace*{-0.5pt} 
\noindent 
The idea is to predict the current signal sample $x_k$ with a
collection of past samples $X_k=(x_{k-d-n}, n = 0, 1,\ldots, N-1)^t$,
the delay $d \geq 1$ being fixed arbitrarily. This is possible, only
if the target sample shares enough information with (i.e., is
sufficiently correlated to) the previous ones. In other words, the
only predictable part of the signal is the one whose correlation
length is sufficiently large (i.e., long-term sinusoids or ringdowns).
On the other hand, the broad band noise cannot be predicted, as it is
not possible to guess the next value in this way.  The prediction
$y_k$ of $x_k$ is obtained through a linear combining of these data
samples weighed by the corresponding coefficients $w_n$, forming the
tap-weight vector $W=(w_n, n = 0, 1,\ldots, N-1)^t$, therefore leading
to $y_k=W^t X_k$.

The optimal tap-weight vector $W^{*}$ which leads to the smallest
prediction error $e_k=y_k-x_k$ in the mean square sense can be proved
to minimize a convex cost function. This minimization can be done
using an approach similar to the steepest descent method with the
following evolution equation for the tap-weights referred to as Least
Mean Square or LMS algorithm~\cite{widrow84:_adapt}~:
\begin{equation}
W_{k+1}=W_{k}+\mu e_k X_k, 
\end{equation}
where the step gain parameter $\mu$ is an adjustable parameter. The
tap-weight coefficients are renewed iteratively so that to converge
and stabilize in a neighborhood of $W^{*}$ whose size is defined by
$\mu$.

Once the filter has converged, we reject the predicted part of the
signal (i.e., $y_k$) corresponding to the long-term sinusoids and the
ringdown noise and we send the rest of the signal (i.e., $e_k$) for
further analysis for detection.

This LMS based prediction method is referred to as \textit{adaptive
  line enhancer (ALE)}~\cite{widrow84:_adapt}. In this context, the
term ``adaptive'' has two different meanings.  First, it means that it
will auto-adjust to reach for the best setup for a problem which is
not initially precisely defined.  Second, it is also able to follow
changes in the characteristics of the data being processed in case
they occur.

Convergence time, frequency tracking ability and frequency resolution
are controlled by the three adjustable parameters~: the number of
tap-weight coefficients $N$, the step gain parameter $\mu$ and the
prediction depth $d$. One can take advantage of certain settings of
the ALE to select a family of signals intead of another.

\newpage
\section{The ALE in practice}
\vspace*{-0.5pt} 
\noindent 
\textbf{Structure of the algorithm} --- We decompose the frequency
axis in $p$ disjoint frequency subbands of the same size. In each
subbands, we apply twice the ALE with different sets of parameters.
In the first step, the adjustable parameters are tuned to best remove
long-term sinusoidal components of the noise ; whereas in the second
one, the target is the short-time oscillatory transients (see~
\cite{chassande-mottin00:_adapt} for details about parameter
adjustment).

The ALE needs to be applied only in the parts of the signal which
appears non-Gaussian. Some refinements are adjunct for this purpose~:
the acceptation (dismissal) of the first cleaning step relies on the
detection of a long-term sinusoids of sufficient amplitude in the
data.  In the second step, the ALE is applied only if the filtered
output $y_k$ deviates from Gaussianity. Details about these additional
vetos can be found in~\cite{chassande-mottin00:_adapt}.

\begin{figure}[htbp]
  \vspace*{13pt}
  \centerline{\psfig{file=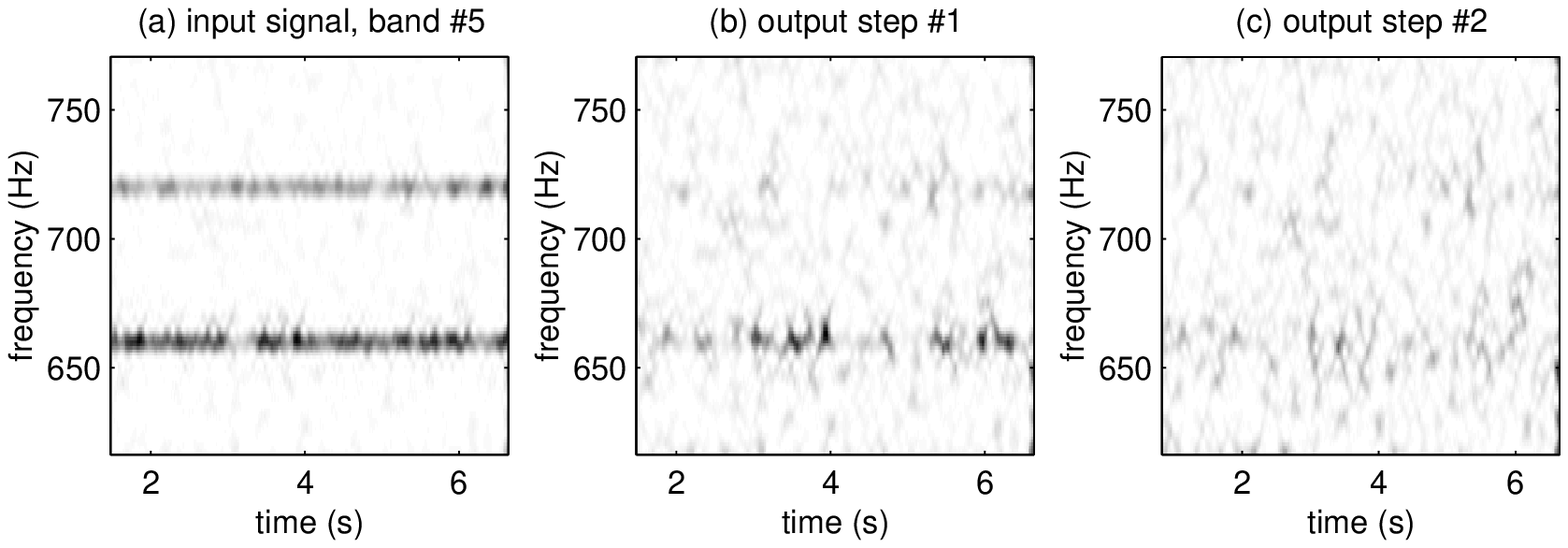,width=11cm}} %100 percent
  \vspace*{13pt} 
    \fcaption{\label{stepfig} \textbf{Illustration of the denoising
      procedure on Caltech proto-type data}. In the subband \#5
    (between 617 Hz and 771 Hz), the signal~\protect\cite{grasp} (the data
    were taken on the October, 14th 1994, frame \#2) contains two power
    line harmonics (at 660 Hz and 720 Hz), as we see on the
    spectrogram (\textbf{a}) (this is a time-frequency representation
    of the signal energy.  Dark regions are associated to large values
    of the energy density).  We apply the ALE a first time to suppress
    long-term components (see spectrogram (\textbf{b})) and a second
    run (\textbf{c}) eliminates artefacts of shorter duration (such as fast
    fluctuations in the harmonic envelope).}
\end{figure}

\begin{figure}[htbp]
  \vspace*{13pt}
  \centerline{\psfig{file=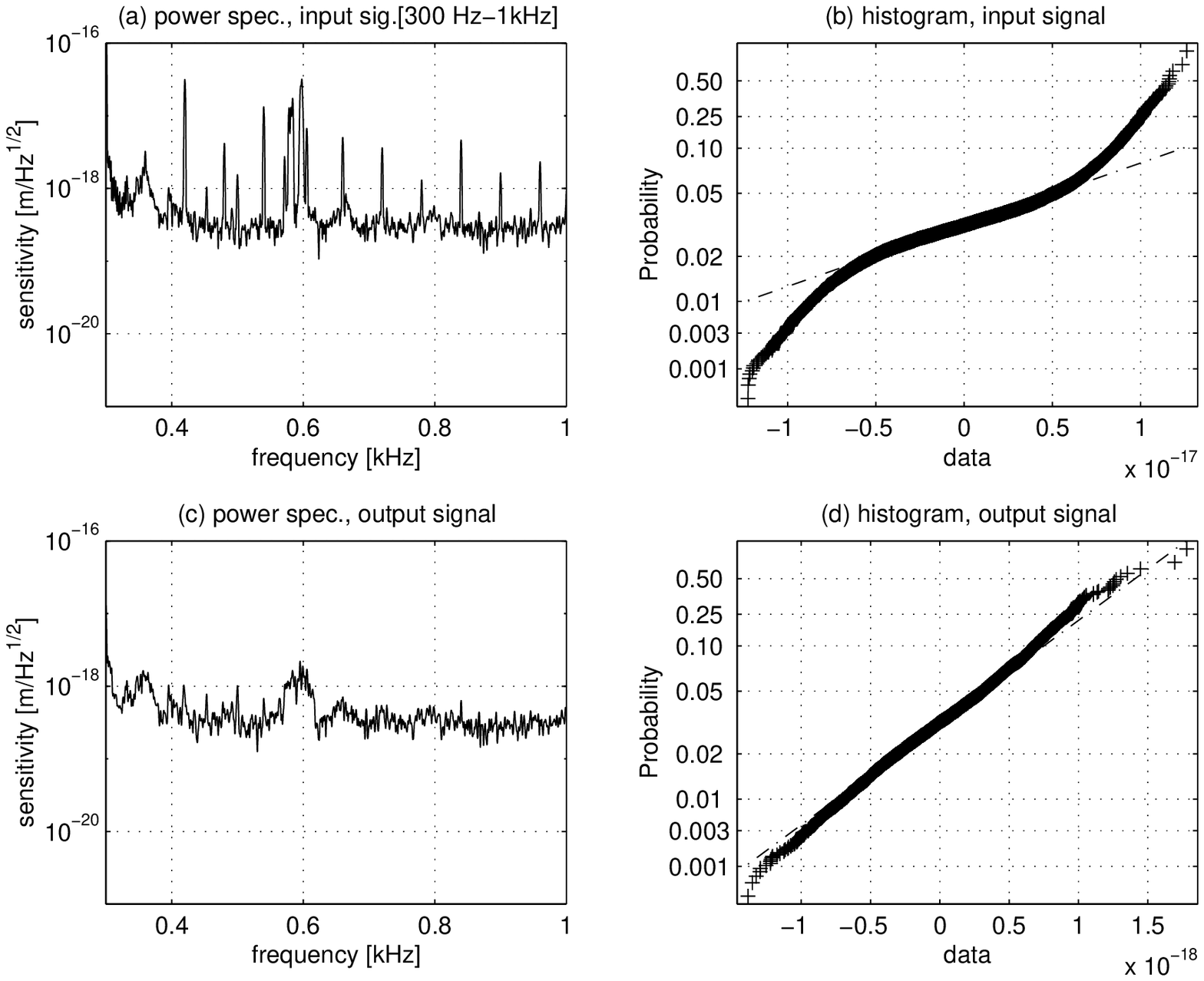,width=11cm}} %100 percent
  \vspace*{13pt} \fcaption{\label{pshist} \textbf{Comparison between
      power spectra and histograms of input/output signals}.
    The figure depicts power spectra (left column) and histograms
    (right column) of the Caltech 40 meter signal (top row) in the
    operating frequency band, between 300 Hz and 1kHz and the same
    signal after denoising (bottom row). The histograms are displayed
    in a graph with special axes where a Gaussian bell curve should
    appeared as a straight line.}
\end{figure}

\vspace*{2mm}
\noindent
\textbf{Results on Caltech 40m proto-type data} --- We have applied
the algorithm to the Caltech 40meter proto-type data taken in October
1994. Figure \ref{stepfig} illustrates how the algorithm is operating
in the fifth frequency subband (from 617 Hz to 771 Hz) among the $p=32$
ones being processed. Figure \ref{pshist} shows comparisons between
the power spectra and histograms of the signal before and after
denoising.

\section{Concluding remarks}
\vspace*{-0.5pt} 
\noindent 
The originality of the proposed approach lies in the fact that it is
possible to treat oscillatory transients. However, it remains that a
comparison of the performances achieved by this algorithm on noise
features of longer duration with other existing methods need to be
done. Finally, although undertaken in~
\cite{chassande-mottin00:_adapt}, this method suffers from the lack of
a complete statistical characterization.

\vspace*{2mm}
\noindent
\textbf{Acknowledgments} --- We would like to thank the LIGO
collaboration for providing us the Caltech 40meter proto-type data.

\nonumsection{References}
%\bibliographystyle{unsrt}
%\bibliography{gwda99}

\end{document}